\input harvmac.tex
\input amssym.tex

\font\eighteenmsb=msbm10 scaled\magstep3
\newfam\bigfam
\textfont\bigfam=\eighteenmsb

\def\bigC{{\fam\bigfam{C}}}
\def\bigP{{\fam\bigfam{P}}}

\def\C{\Bbb{C}}
\def\P{\Bbb{P}}
\def\R{\Bbb{R}}
\def\l{\lambda}
\def\cn{{\cal N}}

\lref\AbbottUM{
  M.~C.~Abbott, I.~Aniceto and O.~O.~Sax,
  ``Dyonic Giant Magnons in $\C \P^3$: Strings and Curves at Finite $J$,''
  arXiv:0903.3365 [hep-th].
}

\lref\HollowoodSC{
  T.~J.~Hollowood and J.~L.~Miramontes,
  ``A New and Elementary $\C \P^n$ Dyonic Magnon,''
  arXiv:0905.2534 [hep-th].
}

\lref\Zakharov{
  V.~E.~Zakharov and A.~V.~Mikhailov,
  ``Relativistically Invariant Two-Dimensional Models In Field Theory
  Integrable By The Inverse Problem Technique. (In Russian),''
  Sov.\ Phys.\ JETP {\bf 47}, 1017 (1978)
  [Zh.\ Eksp.\ Teor.\ Fiz.\  {\bf 74}, 1953 (1978)] $\bullet$
  V.~E.~Zakharov and A.~V.~Mikhailov,
  ``On The Integrability Of Classical Spinor Models In Two-Dimensional
  Space-Time,''
  Commun.\ Math.\ Phys.\  {\bf 74}, 21 (1980).
}

\lref\HarnadWE{
  J.~P.~Harnad, Y.~Saint Aubin and S.~Shnider,
  ``B\"acklund Transformations For Nonlinear Sigma Models With Values In
  Riemannian Symmetric Spaces,''
  Commun.\ Math.\ Phys.\  {\bf 92}, 329 (1984).
}

\lref\HofmanXT{
  D.~M.~Hofman and J.~M.~Maldacena,
  ``Giant magnons,''
  J.\ Phys.\ A  {\bf 39}, 13095 (2006)
  [arXiv:hep-th/0604135].
}

\lref\Dorey{
  N.~Dorey,
  ``Magnon bound states and the AdS/CFT correspondence,''
  J.\ Phys.\ A  {\bf 39}, 13119 (2006)
  [arXiv:hep-th/0604175] $\bullet$
  H.~Y.~Chen, N.~Dorey and K.~Okamura,
  ``Dyonic giant magnons,''
  JHEP {\bf 0609}, 024 (2006)
  [arXiv:hep-th/0605155].
}

\lref\SpradlinWK{
  M.~Spradlin and A.~Volovich,
  ``Dressing the giant magnon,''
  JHEP {\bf 0610}, 012 (2006)
  [arXiv:hep-th/0607009].
}

\lref\KalousiosXY{
  C.~Kalousios, M.~Spradlin and A.~Volovich,
  ``Dressing the giant magnon II,''
  JHEP {\bf 0703}, 020 (2007)
  [arXiv:hep-th/0611033].
}

\lref\IshizekiKH{
  R.~Ishizeki, M.~Kruczenski, M.~Spradlin and A.~Volovich,
  ``Scattering of single spikes,''
  JHEP {\bf 0802}, 009 (2008)
  [arXiv:0710.2300 [hep-th]].
}

\lref\Jevicki{
  A.~Jevicki, C.~Kalousios, M.~Spradlin and A.~Volovich,
  ``Dressing the Giant Gluon,''
  JHEP {\bf 0712}, 047 (2007)
  [arXiv:0708.0818 [hep-th]] $\bullet$
  A.~Jevicki, K.~Jin, C.~Kalousios and A.~Volovich,
  ``Generating AdS String Solutions,''
  JHEP {\bf 0803}, 032 (2008)
  [arXiv:0712.1193 [hep-th]] $\bullet$
  A.~Jevicki and K.~Jin,
  ``Solitons and AdS String Solutions,''
  Int.\ J.\ Mod.\ Phys.\  A {\bf 23}, 2289 (2008)
  [arXiv:0804.0412 [hep-th]].
}

\lref\KalousiosGZ{
  C.~Kalousios, G.~Papathanasiou and A.~Volovich,
  ``Exact solutions for $N$-magnon scattering,''
  JHEP {\bf 0808}, 095 (2008)
  [arXiv:0806.2466 [hep-th]].
}

\lref\AnicetoPC{
  I.~Aniceto and A.~Jevicki,
  ``$N$-body Dynamics of Giant Magnons in $\R\times S^2$,''
  arXiv:0810.4548 [hep-th].
}

\lref\BennaZY{
  M.~Benna, I.~Klebanov, T.~Klose and M.~Smedback,
  ``Superconformal Chern-Simons Theories and AdS${}_4$/CFT${}_3$
  Correspondence,''
  JHEP {\bf 0809}, 072 (2008)
  [arXiv:0806.1519 [hep-th]].
}

\lref\AharonyUG{
  O.~Aharony, O.~Bergman, D.~L.~Jafferis and J.~Maldacena,
  ``$\cn=6$ superconformal Chern-Simons-matter theories, M2-branes and their
  gravity duals,''
  JHEP {\bf 0810}, 091 (2008)
  [arXiv:0806.1218 [hep-th]].
}

\lref\BLG{
  J.~Bagger and N.~Lambert,
  ``Modeling multiple M2's,''
  Phys.\ Rev.\  D {\bf 75}, 045020 (2007)
  [arXiv:hep-th/0611108] $\bullet$
  A.~Gustavsson,
  ``Algebraic structures on parallel M2-branes,''
  Nucl.\ Phys.\  B {\bf 811}, 66 (2009)
  [arXiv:0709.1260 [hep-th]] $\bullet$
  J.~Bagger and N.~Lambert,
  ``Gauge Symmetry and Supersymmetry of Multiple M2-Branes,''
  Phys.\ Rev.\  D {\bf 77}, 065008 (2008)
  [arXiv:0711.0955 [hep-th]] $\bullet$
  J.~Bagger and N.~Lambert,
  ``Comments On Multiple M2-branes,''
  JHEP {\bf 0802}, 105 (2008)
  [arXiv:0712.3738 [hep-th]] $\bullet$
  A.~Gustavsson,
  ``Selfdual strings and loop space Nahm equations,''
  JHEP {\bf 0804}, 083 (2008)
  [arXiv:0802.3456 [hep-th]].
}

\lref\Smatrix{
  N.~Gromov and P.~Vieira,
  ``The all loop AdS${}_4$/CFT${}_3$ Bethe ansatz,''
  JHEP {\bf 0901}, 016 (2009)
  [arXiv:0807.0777 [hep-th]] $\bullet$
  C.~Ahn and R.~I.~Nepomechie,
  ``$\cn=6$ super Chern-Simons theory S-matrix and all-loop Bethe ansatz
  JHEP {\bf 0809}, 010 (2008)
  [arXiv:0807.1924 [hep-th]].
}

\lref\Wilson{
  D.~Berenstein and D.~Trancanelli,
  ``Three-dimensional ${\cal N}=6$ SCFT's and their membrane dynamics,''
  Phys.\ Rev.\  D {\bf 78}, 106009 (2008)
  [arXiv:0808.2503 [hep-th]] $\bullet$
  N.~Drukker, J.~Plefka and D.~Young,
  ``Wilson loops in 3-dimensional $\cn=6$
  supersymmetric Chern-Simons Theory and
  their string theory duals,''
  JHEP {\bf 0811}, 019 (2008)
  [arXiv:0809.2787 [hep-th]] $\bullet$
  B.~Chen and J.~B.~Wu,
  ``Supersymmetric Wilson Loops in $\cn=6$ Super Chern-Simons-matter theory,''
  arXiv:0809.2863 [hep-th] $\bullet$
  J.~Kluson and K.~L.~Panigrahi,
  ``Defects and Wilson Loops in 3d QFT from D-branes in
  AdS${}_4 \times \C\P^3$,''
  arXiv:0809.3355 [hep-th] $\bullet$
  S.~J.~Rey, T.~Suyama and S.~Yamaguchi,
  ``Wilson Loops in Superconformal Chern-Simons Theory and Fundamental Strings
  arXiv:0809.3786 [hep-th] $\bullet$
  C.~S.~Chu and D.~Giataganas,
  ``UV-divergences of Wilson Loops for Gauge/Gravity Duality,''
  JHEP {\bf 0812}, 103 (2008)
  [arXiv:0810.5729 [hep-th]] $\bullet$
  M.~Fujita,
  ``BPS operators from the Wilson loop in the 3-dimensional supersymmetric
  Chern-Simons theory,''
  arXiv:0902.2381 [hep-th].
}

\lref\integrable{
  G.~Arutyunov and S.~Frolov,
  ``Superstrings on AdS${}_4 \times \C\P^3$ as a Coset Sigma-model,''
  JHEP {\bf 0809}, 129 (2008)
  [arXiv:0806.4940 [hep-th]] $\bullet$
  B.~Stefanski~Jr.,
  ``Green-Schwarz action for Type IIA strings on AdS${}_4\times \C\P^3$,''
  Nucl.\ Phys.\  B {\bf 808}, 80 (2009)
  [arXiv:0806.4948 [hep-th]] $\bullet$
  P.~Fr\'e and P.~A.~Grassi,
  ``Pure Spinor Formalism for Osp($N|4$) backgrounds,''
  arXiv:0807.0044 [hep-th] $\bullet$
  G.~Bonelli, P.~A.~Grassi and H.~Safaai,
  ``Exploring Pure Spinor String Theory on AdS${}_4\times \C\P^3$,''
  JHEP {\bf 0810}, 085 (2008)
  [arXiv:0808.1051 [hep-th]] $\bullet$
  J.~Gomis, D.~Sorokin and L.~Wulff,
  ``The complete AdS${}_4 \times \C\P^3$
  superspace for the type IIA superstring and
  D-branes,''
  arXiv:0811.1566 [hep-th].
}

\lref\numeroustests{
  T.~McLoughlin and R.~Roiban,
  ``Spinning strings at one-loop in AdS${}_4 \times \P^3$,''
  JHEP {\bf 0812}, 101 (2008)
  [arXiv:0807.3965 [hep-th]] $\bullet$
  L.~F.~Alday, G.~Arutyunov and D.~Bykov,
  ``Semiclassical
  Quantization of Spinning Strings in AdS${}_4 \times \C\P^3$,''
  JHEP {\bf 0811}, 089 (2008)
  [arXiv:0807.4400 [hep-th]] $\bullet$
  C.~Krishnan,
  ``AdS${}_4$/CFT${}_3$ at One Loop,''
  JHEP {\bf 0809}, 092 (2008)
  [arXiv:0807.4561 [hep-th]] $\bullet$
  T.~McLoughlin, R.~Roiban and A.~A.~Tseytlin,
  ``Quantum spinning strings in AdS${}_4
  \times \C\P^3$: testing the Bethe Ansatz
  proposal,''
  JHEP {\bf 0811}, 069 (2008)
  [arXiv:0809.4038 [hep-th]].
}

\lref\spinchain{
  D.~Gaiotto and X.~Yin,
  ``Notes on superconformal Chern-Simons-matter theories,''
  JHEP {\bf 0708}, 056 (2007)
  [arXiv:0704.3740 [hep-th]] $\bullet$
  J.~A.~Minahan and K.~Zarembo,
  ``The Bethe ansatz for superconformal Chern-Simons,''
  JHEP {\bf 0809}, 040 (2008)
  [arXiv:0806.3951 [hep-th]] $\bullet$
  D.~Bak and S.~J.~Rey,
  ``Integrable Spin Chain in Superconformal Chern-Simons Theory,''
  JHEP {\bf 0810}, 053 (2008)
  [arXiv:0807.2063 [hep-th]] $\bullet$
  B.~I.~Zwiebel,
  ``Two-loop Integrability of Planar $\cn=6$
  Superconformal Chern-Simons Theory,''
  arXiv:0901.0411 [hep-th] $\bullet$
  J.~A.~Minahan, W.~Schulgin and K.~Zarembo,
  ``Two loop integrability for Chern-Simons theories with
  $\cn=6$ supersymmetry,''
  arXiv:0901.1142 [hep-th].
}

\lref\cpthreemagnons{
  D.~Gaiotto, S.~Giombi and X.~Yin,
  ``Spin Chains in $\cn=6$ Superconformal Chern-Simons-Matter Theory,''
  arXiv:0806.4589 [hep-th] $\bullet$
%
  G.~Grignani, T.~Harmark, M.~Orselli and G.~W.~Semenoff,
  ``Finite size Giant Magnons in the string dual of $\cn=6$ superconformal
  Chern-Simons theory,''
  JHEP {\bf 0812}, 008 (2008)
  [arXiv:0807.0205 [hep-th]] $\bullet$
%
  C.~Ahn and P.~Bozhilov,
  ``Finite-size effects of Membranes on AdS${}_4\times S^7$,''
  JHEP {\bf 0808}, 054 (2008)
  [arXiv:0807.0566 [hep-th]] $\bullet$
%
  B.~Chen and J.~B.~Wu,
  ``Semi-classical strings in AdS${}_4 \times \C\P^3$,''
  JHEP {\bf 0809}, 096 (2008)
  [arXiv:0807.0802 [hep-th]] $\bullet$
%
  D.~Astolfi, V.~G.~M.~Puletti, G.~Grignani, T.~Harmark and M.~Orselli,
  ``Finite-size corrections in the $SU(2) \times SU(2)$
  sector of type IIA string
  theory on AdS${}_4 \times \C\P^3$,''
  Nucl.\ Phys.\  B {\bf 810}, 150 (2009)
  [arXiv:0807.1527 [hep-th]] $\bullet$
%
  B.~H.~Lee, K.~L.~Panigrahi and C.~Park,
  ``Spiky Strings on AdS${}_4 \times \C\P^3$,''
  JHEP {\bf 0811}, 066 (2008)
  [arXiv:0807.2559 [hep-th]] $\bullet$
%
  I.~Shenderovich,
  ``Giant magnons in AdS${}_4/CFT_3$: dispersion, quantization and finite--size
  corrections,''
  arXiv:0807.2861 [hep-th] $\bullet$
%
  R.~C.~Rashkov,
  ``A note on the reduction of the AdS${}_4 \times \C\P^3$
  string sigma model,''
  Phys.\ Rev.\  D {\bf 78}, 106012 (2008)
  [arXiv:0808.3057 [hep-th]] $\bullet$
%
  C.~Ahn, P.~Bozhilov and R.~C.~Rashkov,
  ``Neumann-Rosochatius integrable system for strings on AdS${}_4
  \times \C\P^3$,''
  JHEP {\bf 0809}, 017 (2008)
  [arXiv:0807.3134 [hep-th]] $\bullet$
%
  S.~Ryang,
  ``Giant Magnon and Spike Solutions with Two Spins in
  AdS${}_4 \times \C\P^3$,''
  JHEP {\bf 0811}, 084 (2008)
  [arXiv:0809.5106 [hep-th]] $\bullet$
%
  D.~Bombardelli and D.~Fioravanti,
  ``Finite-Size Corrections of the $\C\P^3$ Giant Magnons: the
  L\"{u}scher terms,''
  arXiv:0810.0704 [hep-th] $\bullet$
%
  T.~Lukowski and O.~O.~Sax,
  ``Finite size giant magnons in the $SU(2) \times SU(2)$ sector of
  AdS${}_4 \times \C\P^3$,''
  JHEP {\bf 0812}, 073 (2008)
  [arXiv:0810.1246 [hep-th]] $\bullet$
%
  C.~Ahn and P.~Bozhilov,
  ``Finite-size Effect of the Dyonic Giant Magnons in
  $\cn=6$ super Chern-Simons
  Theory,''
  arXiv:0810.2079 [hep-th] $\bullet$
%
  B.~H.~Lee and C.~Park,
  ``Unbounded Multi Magnon and Spike,''
  arXiv:0812.2727 [hep-th].
}

\lref\Abbott{
  M.~C.~Abbott and I.~Aniceto,
  ``Giant Magnons in AdS${}_4 \times \C\P^3$:
  Embeddings, Charges and a Hamiltonian,''
  arXiv:0811.2423 [hep-th].
}

\lref\GrignaniIS{
  G.~Grignani, T.~Harmark and M.~Orselli,
  ``The $SU(2) \times SU(2)$ sector in the string dual of $\cn=6$
  superconformal
  Chern-Simons theory,''
  Nucl.\ Phys.\  B {\bf 810}, 115 (2009)
  [arXiv:0806.4959 [hep-th]].
}

\lref\HollowoodTW{
  T.~J.~Hollowood and J.~L.~Miramontes,
  ``Magnons, their Solitonic Avatars and the Pohlmeyer Reduction,''
  arXiv:0902.2405 [hep-th].
}

\lref\SuzukiSC{
  R.~Suzuki,
  ``Giant Magnons on $\C\P^3$ by Dressing Method,''
  arXiv:0902.3368 [hep-th].
}

\Title
{\vbox{
\baselineskip12pt
\hbox{Brown-HET-1575}
}}
{\vbox{
\centerline{Dressed Giant Magnons on $\bigC\bigP^3$}
}}

\centerline{
Chrysostomos Kalousios,
Marcus Spradlin
and Anastasia Volovich
}

\vskip .5in
\bigskip
\centerline{Brown University}
\centerline{Providence, Rhode Island 02912 USA}

\vskip .5in

\centerline{\bf Abstract}

A new example of AdS/CFT duality
relating IIA string theory on AdS${}_4 \times \C \P^3$
to $\cn = 6$ superconformal
Chern-Simons theory has
recently been provided by ABJM.
By now a number of papers have considered
particular giant magnon classical
string solutions in the $\C\P^3$ background, corresponding
to excitations in the spin chain picture of the dual field theory.
In this paper
we apply the $\C\P^3 = SU(4)/S(U(3) \times U(1))$
dressing method to the
problem of constructing general classical string solutions describing
various
configurations of giant magnons.
As a particular application we present a new giant
magnon solution on~$\C\P^3$.  Interestingly the dressed solution
carries only a single $SO(6)$ charge, in contrast with the dyonic
magnons found in previous applications of the dressing method.

\Date{}

\listtoc
\writetoc

\newsec{Introduction}

Motivated by the work of Bagger, Lambert
and Gustavsson~\BLG\ on maximally
superconformal field theories in three dimensions,
Aharony, Bergman, Jafferis, and Maldacena (ABJM)
constructed~\AharonyUG\ an $\cn=6$ superconformal Chern-Simons theory with
$U(N) \times U(N)$ gauge symmetry at levels $(k,-k)$
that is believed to be dual
to $M$-theory on AdS${}_4 \times S^7/Z_k$ (see also~\BennaZY).
ABJM
further  considered the $N,k \rightarrow \infty$ limit keeping the 't
Hooft coupling $\l=N/k$ fixed and
conjectured that in this limit
the $\cn=6$ field theory is dual to type IIA string
theory on AdS${}_4 \times \C\P^3$.

Given the important role
that integrability has played in exploring the structure
of ${\cal N} = 4$ Yang-Mills theory and its dual, it is natural that
this new example of AdS/CFT provides an arena for
further studying aspects of integrability in gauge/string duality.
The worldsheet theory for IIA strings on AdS${}_4 \times \C\P^3$ has
been constructed and its possible integrability explored in~\integrable,
while
on the Chern-Simons
side the anomalous dimensions of local operators are
apparently encoded in
integrable spin chain Hamiltonian~\spinchain.
An exact magnon S-matrix for this spin chain has been
proposed
in~\Smatrix, numerous tests of these
proposals have been carried out in~\numeroustests, and
aspects of
Wilson loops have been studied in~\Wilson.

Hofman and Maldacena~\HofmanXT\ identified
the string theory dual of an elementary magnon in the spin chain
description of $\cn=4$ Yang-Mills theory as a
a particular classical
open string configuration on an $\R \times S^2$ subset of
AdS${}_5 \times S^5$, called the giant magnon.  The
study of
giant magnons
and their BPS bound states~\Dorey\ has
provided a wealth of detailed information about AdS/CFT.
Naturally therefore a number of
papers~\refs{\cpthreemagnons,\GrignaniIS,\Abbott,\HollowoodTW}
have explored
in detail the properties of various giant magnon solutions relevant
to the ABJM incarnation of AdS${}_4$/CFT${}_3$.

The dressing method of Zakharov and Mikhailov~\Zakharov\ provides
an algorithm
to directly construct  solutions of classically integrable
equations.
This method has proven  useful for the construction of various
giant magnon solutions, including magnons on
spheres~\refs{\SpradlinWK,\KalousiosXY,\IshizekiKH}
and on anti-de Sitter space~\Jevicki.
An explicit solution describing the scattering of $N$
giant magnons on $\R \times S^3$ was also presented
in~\KalousiosGZ, and their dynamics on $S^2$ were studied
in~\AnicetoPC.
Since the equations of motion for a string on $\R \times \C\P^3$ are
also classically integrable, these techniques
can be employed here as well.
In this paper we demonstrate the application
of the dressing method for $SU(4)/S(U(3) \times U(1))$ coset
model (due to Harnad et.~al.~\HarnadWE)  to the problem of
constructing $\C\P^3$ giant magnon solutions.

An important feature of the dressing method,
which has been exploited for example in~\refs{\SpradlinWK,\KalousiosXY,
\IshizekiKH,\KalousiosGZ},
is that
repeated application can be used
to generate explicit classical string solutions describing
the scattering of any number of giant magnons (or, when applicable,
bound states thereof).
In complex sigma models
each application of the dressing method
introduces two new non-trivial parameters into the solution, which
in familiar cases~\refs{\SpradlinWK,\KalousiosXY} specify the magnon's
momentum $p$ and the value of a second $SO(6)$ charge $J_2$
which it carries in addition
to the charge $J$ carried by the ground state in an orthogonal plane.
Such dressed solutions are therefore dyonic giant magnons of the
type introduced by Dorey~\Dorey.
However in the present application of the dressing method we find
something of a surprise---the dressed solution
given below in (4.6) indeed has two
parameters, but it carries only a single $SO(6)$ charge $J$, having $J_2 = 0$.
The dressing method apparently finds
a one-parameter family of solutions for any fixed value of
the charges $p$ and $J$, but no dyonic magnons.
We do not present any multi-magnon
solutions here, but the algebra involved is no more complicated
than
for the solutions studied in~\refs{\KalousiosXY,\KalousiosGZ}.

\bigskip
\noindent
{\bf Note Added.}
The results in this note were obtained some time ago but were delayed
due to the disappointment of not being able to find more
general (two-charge) solutions.  We were motivated to publish our
results by the recent appearance of~\HollowoodTW\ which includes,
in addition to a formula (7.15) for
the new $\C\P^3$ magnon that is identical to our solution (4.6),
a very thorough analysis demonstrating that the dressing method by itself
cannot produce any such ``dyonic'' solutions.  This conclusion
has also been reached in~\SuzukiSC.
Elementary dyonic giant magnon solutions on $\C \P^3$ have finally
been obtained in~\AbbottUM\ for a special value of $p$ and
in~\HollowoodSC\ generally.
The solution (4.6) below is evidently a neutral composite of two such
elementary magnons~\HollowoodSC.

\newsec{The $\C\P^3$ Model}

The
$\C\P^3$ model may be described by a complex four-component vector $n$ with
lagrangian density
\eqn\action{
2 \sqrt{2 \lambda} \,{\cal L} =
- \partial_\mu n^\dagger \cdot \partial^\mu n
+ (n^\dagger \cdot \partial_\mu n) (\partial^\mu n^\dagger\cdot  n)
- \Lambda(n^\dagger\cdot n - 1)
}
where $\lambda=N/k$ is the 't Hooft coupling.
The Lagrange multiplier constrains the fields to lie on $S^7 \in \C^4$, while
the local $U(1)$ invariance of~\action\ allows us to identify $n \sim e^{i \Lambda(x)} n$,
thereby reducing the configuration space to $S^7/U(1) = \C\P^3$.
The action possesses an $SU(4)$ symmetry with Noether currents
\eqn\currents{
J_\mu^a =2\sqrt{2\lambda} \,{\rm Im}
[(n^\dagger \cdot T^a \partial_\mu n)
- (n^\dagger \cdot T^a n)
(n^\dagger \cdot  \partial_\mu n)],
}
where $T^a$ are generators of $SU(4)$.
The equations of motion (after eliminating the Lagrange multiplier) are
\eqn\eom{
- \partial^2 n + (n^\dagger \cdot \partial^2 n) n
+ 2 (n^\dagger \cdot \partial_\mu n) \partial^\mu n
+ 2 (\partial_\mu n^\dagger \cdot n)(n^\dagger \cdot \partial^\mu n) n
= 0.
}
To
describe
classical strings on $\R \times \C\P^3$
(with a trivial time coordinate), the
equations of motion must be supplemented with the Virasoro constraints
\eqn\vir{\eqalign{
(\partial_+ n^\dagger \cdot \partial_+ n)
- (n^\dagger \cdot \partial_+ n)
(\partial_+ n^\dagger \cdot n) &= {1 \over 4}, \cr
(\partial_- n^\dagger \cdot \partial_- n) -
(n^\dagger \cdot \partial_- n) (\partial_- n^\dagger\cdot n) &= {1 \over 4},
}}
where we have used light-cone coordinates
$x_+={1 \over 2}(x-t),~x_-={1 \over 2}(x+t)$ and the derivatives
are with respect to those coordinates,
$\partial_+=\partial_x-\partial_t$, $\partial_-=\partial_x+\partial_t$.

Several classes of solutions to
the equation of motion~\eom\ and
the Virasoro constraints~\vir\ may be obtained
by embedding known giant magnon solutions that live on $S^2$
or $S^3$ into $\C\P^3$
(an extensive discussion of these embeddings has been
given in~\Abbott).
As a first example, let $(X_1,X_2,X_3)$ be
coordinates satisfying $X_1^2+X_2^2+X_3^2 = 1$.
An isometric embedding $S^2 \to \C\P^3$ is given by
\eqn\utut{
n^{\rm T} = {1 \over \sqrt{2(1-X_3)}}\pmatrix{
{ X_1 + i X_2} & 1 - X_3 & 0& 0}.
}
In this manner any solution $X^i= (X_1,X_2,X_3)$ of
string theory on $\R \times S^2$
\eqn\aaa{\eqalign{
- \partial^2 X^i + (X \cdot \partial^2 X) X^i &= 0,\cr
\partial_+ X \cdot \partial_+ X = \partial_- X \cdot \partial_- X &= 1,}
}
lifts to a solution of~\eom\ and~\vir.
In general it may be necessary to rescale the worldsheet coordinates
$x,t$ in order to satisfy~\vir\ with the normalization shown.  Such
a rescaling does not affect the equations of motion~\eom.
Similarly we can consider
what has been called the
``$S^2 \times S^2$'' embedding
in the literature. This is given by the map
\eqn\aaa{
n^{\rm T} = {1 \over 2\sqrt{(1-X_3)}}\pmatrix{
X_1 + i X_2& 1 - X_3& X_1+i X_2& 1-X_3}
}
whose image inside $\C\P^3$ is actually~\Abbott\ just a single $S^2$ partially
rotated into two orthogonal directions compared to~\utut.

In the case of magnons living on $S^3$ we can parameterize the
unit 3-sphere with embedding coordinates $X^i= (X_1,X_2,X_3,X_4)$.
Given any such solution $X^i$ describing a classical string
on $\R \times S^3$ there are two possible natural embeddings
into a solution of the $\C\P^3$ equations, given alternately by
\eqn\aaa{
n^{\rm T}=\pmatrix{X_1 & X_2 & X_3 & X_4}
}
or
\eqn\aaa{
n^{\rm T}={1\over \sqrt{2}}
\pmatrix{X_1+i X_2 & X_3+i X_4 & X_1-i X_2 & X_3-i X_4},
}
whose images are both $\R\P^3 \subset \C\P^3$~\Abbott.

To provide a concrete example
we remind the reader of the solution describing
Dorey's dyonic magnon~\Dorey\ on $S^3$,
\eqn\aaa{\eqalign{
X^1+iX^2&=e^{it/2}(\cos{\textstyle{p\over 2}}+
i\sin{\textstyle{p\over 2}}\tanh{\textstyle{u \over 2}}),\cr
X^3+iX^4&=e^{i {v}/2 }
\sin{\textstyle{p\over 2}}\,{\rm sech}\,{\textstyle{u \over 2}},
}}
where
\eqn\aaa{
{u}={i}(Z(\lambda_1)-Z(\bar{\lambda}_1)),
\qquad {v}= Z(\lambda_1)+Z(\bar{\lambda}_1)-t
}
in terms of
\eqn\aaa{
Z(\lambda)={x_{+} \over \lambda-1}+{x_{-} \over \lambda +1}.
}
The resulting $\C\P^3$ solution involves
putting two conjugate (and hence, oppositely charged)
dyonic giant magnons together.
The scaling of the world sheet coordinates $(x,t)$ by 1/2
compared to~\Dorey\ leaves the
equation of motion~\eom\ intact but is necessary in order to preserve
the normalization of the
Virasoro constraints~\vir.  In the above we have used the parameterization
$\l_1=r e^{ip/2}$ of the spectral parameter, where $p$ is the momentum of the
magnon and $r$ is related to its charge.  In the limit $r\rightarrow 1$ we
recover the ``$S^2 \times S^2$'' solution presented in~\GrignaniIS.

More generally we can take any known giant magnon solution on $S^3$
(such as the general $N$-magnon solution found in~\KalousiosGZ) and embed it
into the $\C\P^3$ model, thus obtaining a new classing string solution
moving on $\R \times \C\P^3$ (again, it may be necessary to also scale
the worldsheet coordinates to preserve the normalization given
in~\vir).
Besides these `trivial' solutions reviewed here, the $\C\P^3$ model admits
more general solutions that can be obtained via the dressing method, to which
we now turn our attention.

\newsec{The Dressing Method for the $\C\P^3$ Coset}

In order to apply the dressing method as outlined in \HarnadWE\ we first embed
the $\C\P^3$ vector field $n$ into an $SU(4)$ principal chiral field $g$.
This may be done by noting that
\eqn\aaa{
\C\P^3 = \{ g \in SU(4) : g \Omega g \Omega = 1\}, \qquad
{\rm with}~
\Omega = \pmatrix{+1 & 0 & 0 & 0 \cr
0 & -1 & 0 & 0 \cr
0 & 0 & -1 & 0 \cr
0 & 0 & 0 & - 1}.
}
In order to understand this embedding, first
observe that if $g$ satisfies
$g^\dagger g = 1$ and
$g \Omega g \Omega = 1$
then the matrix
\eqn\aaa{
P = {1 + \Omega g \over 2}
}
is a hermitian projection operator.
Since $\det g = 1$ it follows that $\det (2 P - 1) = - 1$
so the rank of $P$ must be either 1 or 3.
In fact we can without loss of generality take $P$ to have rank 1 since
otherwise we could just replace $P \to 1 - P$ throughout this analysis.
Then we identify the vector $n$ as the (unit-normalized)
image of $P$.

Conversely, given a unit vector $n$ we take
\eqn\aaa{
g = \Omega(2 P - 1) \qquad {\rm with}~P = n n^\dagger,
}
which is easily seen to satisfy $g \Omega g \Omega = 1$ and
$g \in SU(4)$.

Under this embedding, the lagrangian \action~becomes proportional to
\eqn\aaa{
{\cal L}=\Tr[(g^{-1} \partial_{\mu} g)^2],
}
the equation of motion~\eom\ becomes equivalent to
the principal chiral model equation
\eqn\aaa{
\partial_+ \partial_- g - {1 \over 2} (\partial_+ g g^{-1} \partial_- g
+ \partial_- g g^{-1} \partial_+ g) = 0,
}
while the Virasoro constraints~\vir\ map into
\eqn\aaa{
\Tr[(g^{-1} \partial_+ g)^2] = -2, \qquad
\Tr[(g^{-1} \partial_- g)^2] = -2.
}

Next we recall Theorem 4.2 of~\HarnadWE.  Given any solution $g$ of
the $SU(4)$ principal chiral model which
satisfies $g \Omega g \Omega = 1$, we first solve the auxiliary
system
\eqn\auxsys{
\partial_+ \Psi = {\partial_+ g g^{-1} \Psi \over 1 - \lambda}, \qquad
\partial_- \Psi = {\partial_- g g^{-1} \Psi \over 1 + \lambda}
}
to find $\Psi(\lambda)$ as a function of the auxiliary complex
parameter $\lambda$, subject to the initial condition
\eqn\aaa{
\Psi(0) = g,
}
the $SU(4)$ constraints
\eqn\aaa{
\det \Psi(0) = 1, \qquad [\Psi(\bar{\lambda})]^\dagger
\Psi(\lambda) = 1,
}
as well as the coset constraint
\eqn\aaa{
\Psi(\lambda) = \Psi(0) \Omega \Psi(1/\lambda) \Omega.
}

With $\Psi(\lambda)$ in hand a new dressed solution to the coset
model may be constructed algebraically.
The input to
specify a new solution is an arbitrary complex parameter $\lambda_1$
and an arbitrary complex four-vector $e$.
In terms of this data the dressed solution is
$g' = \Psi'(0)$ where
\eqn\aaa{
\Psi'(\lambda) =
\left[ 1 + {Q_1 \over \lambda - \lambda_1}
+ {Q_2 \over \lambda - 1/\lambda_1} \right] \Psi(\lambda)
}
in terms of two matrices $Q_i = X_i F^\dagger_i$ specified by
\eqn\aaa{
F_1 = \Psi(\bar{\lambda}_1) e, \qquad
F_2 = \Psi(0) \Omega \Psi(\bar{\lambda}_1) e
}
and the $X_i$ are the solutions to
\eqn\aaa{
\eqalign{
X_1 { F_1^\dagger F_1 \over \lambda_1 - \bar{\lambda}_1 }
+ X_2 { F_2^\dagger F_1 \over 1/\lambda_1 - \bar{\lambda}_1} &= F_1, \cr
X_1 { F_1^\dagger F_2 \over \lambda_1 - 1/\bar{\lambda}_1} +
X_2 { F_2^\dagger F_2 \over 1/\lambda_1 - 1/\bar{\lambda}_1} &= F_2.
}}

\newsec{Giant Magnon Solutions on $\C\P^3$}

In order to obtain new giant magnon solutions on $\C\P^3$ via the dressing
method described in the previous paragraph we first choose as the vacuum
\eqn\firstvacuum{
n^{\rm T} = \pmatrix{ \cos (t/2) & \sin (t/2) & 0 & 0 }.
}
(a perhaps more obvious, but less useful, choice will be considered below).
The scaling of the worldsheet time coordinate by 2 is necessary if we want our vacuum to satisfy the Virasoro constraints~\vir.  Then the solution to the auxiliary system \auxsys~that satisfies the initial condition and constraints is
\eqn\aaa{
\Psi(\lambda) = \pmatrix{
\cos Z(\lambda) & \sin Z(\lambda)  & 0 & 0 \cr
-\sin Z(\lambda) & \cos Z(\lambda) & 0 & 0 \cr
0 & 0 & 1 & 0 \cr
0 & 0 & 0 & 1 },
}
where
\eqn\aaa{
Z(\lambda)={x_{+} \over \lambda-1}+{x_{-} \over \lambda +1}.
}
Choosing (arbitrarily) the polarization vector
\eqn\polarization{
e^{\rm T}=\pmatrix{1 & 0 & i & 0}.
}
we find the solution
\eqn\firstsolution{
n^{\rm T}={1 \over \sqrt{R}} \pmatrix{n^1 & n^2 & n^3 & n^4},
}
specified by
\eqn\newsolution{\eqalign{
n^1 & = + 2  (1 - \lambda_1^2) \bar{\lambda}_1 \cos (t/2)
+ (1 - |\lambda_1|^2) (\lambda_1 \cos (t/2 -  i u)
+ \bar{\lambda}_1 \cos(t/2 +  i u)) \cr &\qquad
+(\lambda_1 - \bar{\lambda}_1) (\cos(t/2 - v)
+ |\lambda_1|^2 \cos(t/2 + v)),\cr
n^2 & = - 2  (1 - \lambda_1^2) \bar{\lambda}_1 \sin (t/2)
- (1 - |\lambda_1|^2) (\bar{\lambda}_1 \sin(t/2 +  i u) +
\lambda_1 \sin(t/2 -  i u)) \cr & \qquad
- (\lambda_1 - \bar{\lambda}_1) (\sin(t/2 -  v)
+ |\lambda_1|^2 \sin(t/2 +  v)),\cr
n^3  & =  - 2  i (\lambda_1 - \bar{\lambda}_1) (1 - |\lambda_1|^2) \cosh(u/2 + i v/2),\cr
n^4  & =0.
}}
In the above the normalization factor $R$ is given by
\eqn\aaa{R=\sum_{i=1}^4 \bar{n}^i n^i
}
and
\eqn\uvdef{
u=i(Z(\lambda_1)-Z(\bar{\lambda}_1)),
\qquad v=Z(\lambda_1)+Z(\bar{\lambda}_1)-t.
}
The solution~\newsolution\ is identical to the one
presented recently in~\HollowoodTW\ (including, coincidentally,
an almost identical choice of polarization vector~\polarization).

Upon parameterizing $\lambda_1 = r e^{i p/2}$ and evaluating
the Noether charges using~\currents\ we find that
the solution~\firstsolution\ carries only a single nonzero charge $J$
satisfying the dispersion relation
(adapted to the conventions used in~\AbbottUM)
\eqn\disper{
\Delta-{1 \over 2} J= 2 \sqrt{2 \lambda} {1+r^2 \over 2r}
\left|\sin {p \over 2} \right|.
}
As usual for giant magnons, the charge $J$ is itself infinite but
the excitation energy $\Delta - J$ of the magnon above the ground
state (a pointlike string moving at the speed of light) is finite.
Remarkably the formula~\disper\ is identical to the corresponding
one for Dorey's dyonic giant magnon~\Dorey, but the
solution~\newsolution\ apparently carries only a single macroscopic
charge and is hence not ``dyonic'' at all.
The solution does reduce to the original Hofman-Maldacena
magnon~\HofmanXT\ with momentum $p$ when the parameter
$r$, whose physical interpretation at this point is mysterious, is taken to 1.

A second possible choice of vacuum is
\eqn\aaa{
n^{\rm T} = {1 \over \sqrt{2}} \pmatrix{ e^{i t} & 1 & 0 & 0 },
}
which differs from~\firstvacuum\ by an $SU(4)$ rotation
which, importantly, does not commute with $\Omega$.
In this case
the solution to the linear system~\auxsys\ is
\eqn\aaa{
\Psi(\lambda) = {1 \over \sqrt{1 + \lambda}} \pmatrix{
\sqrt{\lambda}\, e^{+ i Z(\lambda)} & + e^{+ i Z(\lambda)} & 0 & 0 \cr
- e^{-i Z(\lambda)} & \sqrt{\lambda}\, e^{-i Z(\lambda)} & 0 & 0 \cr
0 & 0 & \sqrt{1 + \lambda} & 0 \cr
0 & 0 & 0 & \sqrt{1 + \lambda} }.
}
If we choose as polarization vector
\eqn\aaa{
e^{\rm T}=\pmatrix{1 & i & 0 & 0}
}
we find the solution
\eqn\secondsolution{\eqalign{
n^1&=e^{+it/2}\left((\l_1-\bar{\l}_1)|\l_1|^2e^{iv}-(\l_1-\bar{\l}_1)e^{-iv}-i(1-|\l_1|^2)\l_1e^u -i(1-|\l_1|^2)\bar{\l}_1e^{-u}\right),\cr
n^2&=e^{-it/2}\left((\l_1-\bar{\l}_1)|\l_1|^2 e^{-iv}-(\l_1-\bar{\l}_1)e^{iv} +i(1-|\l_1|^2)\l_1 e^{-u}+i(1-|\l_1|^2)\bar{\l}_1e^u\right),\cr
n^3&=n^4=0,
}}
in terms $u$ and $v$ as before in~\uvdef.
(This solution
must of course also be normalized to unit length as in~\firstsolution.)
Remarkably this solution is identical to the bound state of two
Hofman-Maldacena magnons on $S^2$ found in (5.14) of~\SpradlinWK,
embedded into $\C \P^3$ via~\utut.

\bigskip
\noindent
{\bf Acknowledgements}

We have benefited from from correspondence with T.Hollowood
and discussions
with
M. Abbott, I. Aniceto,
N. Bobev, A. Jevicki and G. Papathanasiou.
We are especially grateful to M. Abbott, I. Aniceto and K. Jin for
pointing out an important misstatement in a draft of this paper.
C. K. is grateful to the Institut d'\'Etudes Scientifiques de Carg\`ese,
M. S. to the Banff International Research Station,
and A. V. to the ETH Z\"urich,
for hospitality and support during the course of this work.
This work was supported in part by the
US Department of Energy under contract
DE-FG02-91ER40688 (M. S. (OJI) and A. V.), and the
US National Science Foundation under grants PHY-0638520 (M. S.) and
PHY-0643150
CAREER and PECASE (A. V.).

\listrefs
\end